\documentclass[aip,apl,amsmath,amssymb,reprint]{revtex4-1}
\usepackage[T1]{fontenc}       % Use modern font encodings
\usepackage[utf8]{inputenc}
\usepackage[colorlinks=true,allcolors=blue]{hyperref}
\usepackage{graphicx}

\begin{document}

%\preprint{}

% Title of paper
\title{Passivation of dangling bonds on hydrogenated Si(100)-2$\times$1: a possible method for error correction in hydrogen lithography}
% Error correction for hydrogen lithography on hydrogen-passivated Si(100)-2$\times$1 surfaces



\author{Niko Pavliček}
\email{pav@zurich.ibm.com}
%\thanks{}
\affiliation{IBM Research–Zurich, Säumerstrasse 4, 8803 Rüschlikon, Switzerland}
\author{Zsolt Majzik}
\affiliation{IBM Research–Zurich, Säumerstrasse 4, 8803 Rüschlikon, Switzerland}
\author{Gerhard Meyer}
\affiliation{IBM Research–Zurich, Säumerstrasse 4, 8803 Rüschlikon, Switzerland}
\author{Leo Gross}
\affiliation{IBM Research–Zurich, Säumerstrasse 4, 8803 Rüschlikon, Switzerland}


\date{\today}

\begin{abstract}
Using combined low temperature scanning tunneling microscopy (STM) and atomic force microscopy (AFM), we demonstrate hydrogen passivation of individual, selected dangling bonds (DBs) on a hydrogen-passivated Si(100)-2$\times$1 surface (H--Si) by atom manipulation. This method allows erasing of DBs and thus provides an error-correction scheme for hydrogen lithography. Si-terminated tips (Si tips) for hydrogen desorption and H-terminated tips (H tips) for hydrogen passivation are both created by deliberate contact to the H--Si surface and are assigned by their characteristic contrast in AFM. DB passivation is achieved by transferring the H atom that is at the apex of an H tip to the DB, reestablishing a locally defect-free H--Si surface.
%The tips, the surface and the H passivation by atom manipulation are characterized %using AFM in conjunction with  at $5\,\text{K}$. 
\end{abstract}

\pacs{61.43.-j,68.35.Dv,68.37.Ef,68.37.Ps,71.55.Ak,72.20.Jv,81.65.Rv}% insert suggested PACS numbers in braces on next line

% 61.43.-j Bonds, dangling
% 68.35.Dv Defects, crystal, solid surfaces and interfaces
% 68.37.Ef STM, in study of surface structure
% 68.37.Ps AFM, in surface structure determination
% 71.55.Ak Semiconductors, elemental, impurity and defect levels in
% 72.20.Jv Charge carriers, semiconductors and insulators
% 81.65.Rv Passivation, surface treatment

\maketitle %\maketitle must follow title, authors, abstract and \pacs

\section{Introduction}
%\label{}

Hydrogen lithography \cite{Lyding1994,Shen1995}, that is controlled dehydrogenation of H-passivated semiconductor surfaces has become an important tool in the development of novel atomic scale logic devices and for the study of nanoscale and quantum physics. To this end, individual H atoms are deliberately desorbed from an H-passivated surface using a scanning probe microscopy tip by applying a voltage between tip and sample. An individual DB is created where a H atom is removed. Such DBs can be employed directly to construct atomically precise DB structures or they can be used for further processing. Importantly, DBs provide adsorption sites for single molecules, e.\,g. phosphine, which after incorporation and activation provides individual phosphorous dopants \cite{Schofield2003,Ruess2004}. In addition, individual organic molecules can be captured and studied at individual DBs \cite{Hersam2000,Piva2005,Godlewski2013}. 

Several proposals for novel, disruptive technologies are based on hydrogen lithography with atomic precision. These proposals include silicon-based solid state quantum computers \cite{Simmons2003}, field-coupled silicon atomic quantum dots \cite{Wolkow2014}, quantum Hamiltonian boolean logic gates \cite{Kolmer2015}. Single atom transistors \cite{Fuechsle2012} and atomic-scale wires \cite{Weber2012} fabricated by hydrogen lithography followed by phosphor doping have already been demonstrated. Moreover, DBs provide an interesting model system for studying quantum physics, e.\,g. artificial molecules can be created and studied \cite{Schofield2013,Wood2016}. Importantly, the DBs can adopt different charge states, depending on the global doping concentration and precise location of nearby charges, providing possible electrostatic coupling and switching schemes \cite{Haider2009,Labidi2015}. In addition, the position of a DB can be switched within a dimer \cite{Bellec2013,Engelund2016}, molecules can be adsorbed at DBs \cite{Hersam2000,Piva2005},  and their orientation can be changed by atomic manipulation \cite{Godlewski2016}  representing mechanical atomic switches.

For complex devices, single DBs have to be placed with atomic precision with a high yield. The yield has been extensively studied and continuously been improved over the last two decades \cite{Shen1995,Stokbro1998,Soukiassian2003,Randall2009,Ballard2013,Kolmer2014,Wolkow2014,Moller2017}, now reaching a yield of 0.8 to 0.9 (Ref.~\onlinecite{Wolkow2014}). However, for complex, atomically precise structures needed for the envisioned applications an almost perfect yield is desired. Here, we demonstrate error correction for hydrogen lithography. We describe and demonstrate a method that can be used to cure, i.\,e., passivate individual selected DBs at will, recovering locally a defect-free hydrogen passivated surface. The method is fully compatible with hydrogen lithography. Therefore, faulty written DB structures can be erased and rewritten until all errors are removed yielding atomically precise DB structures.

To the best of our knowledge, erasing single DBs selectively by hydrogenation has not been shown before. Reversible passivation of silicon dangling bonds was demonstrated using a radical organic molecule adsorbed at the DB \cite{Pitters2003}. Hydride formation by atomic manipulation was to date only reported on bare, non-passivated Si surfaces and only from molecular hydrogen (H$_{2}$) by applying voltage pulses \cite{Kuramochi1994,Kuramochi1997,Labidi2012}.

The key for the correction of DB structures is the control of the tip functionalization. It has been shown before that the final tip atom is of utmost importance for AFM imaging with atomic resolution \cite{Gross2009b,Sharp2012,Yurtsever2013,Sweetman2016,Labidi2017} and the importance of the tip for hydrogen lithography has been pointed out \cite{Moller2017}. On hydrogenated Si surfaces two distinctly different tip functionalizations have been achieved, characterized and assigned using the combination of STM and AFM \cite{Sharp2012,Yurtsever2013,Sweetman2016,Labidi2017}. On the one hand a chemically inert tip was assigned as an H tip \cite{Sharp2012,Labidi2017}, and on the other hand a highly reactive tip was assigned as a Si tip \cite{Sweetman2016}, the latter enabling even room-temperature atomic-resolution imaging of organic molecules by AFM \cite{Iwata2015}. We achieved writing and correcting Si DBs by employing these two different tip functionalizations: a Si tip is used to write DBs, that is for hydrogen desorption, and a H tip is used for error correction, that is for hydrogen passivation. In the latter case the H from the tip apex is transferred into the DB, removing the defect, resulting in an H-terminated Si surface that cannot be distinguished from the pristine H--Si surface and at the same time the tip is changed from an H to a Si tip.

\section{Experimental details}
We used a Si(100) wafer with high n-type doping (phosphor dopant concentration of $2\times10^{19}\,\text{cm}^{-3}$, room temperature resistivity of $2-3\,\text{m}\Omega\,\text{cm}$). To prepare the sample in UHV (base pressure $p < 10^{-10}\,\text{mbar}$) the wafer was flash-annealed to $1340\,\text{K}$ several times to remove the native oxide. During the final anneal the chamber pressure remained below $2 \times 10^{-10}\,\text{mbar}$. Subsequently, the sample was H-passivated for $240\,\text{s}$ using a hydrogen atom beam source at a partial H$_{2}$ pressure of $1.3 \times 10^{-7}\,\text{mbar}$, with the sample held at a temperature of $560\,\text{K}$. Then, the sample was transferred to the STM/AFM stage.

\begin{figure}
	\includegraphics{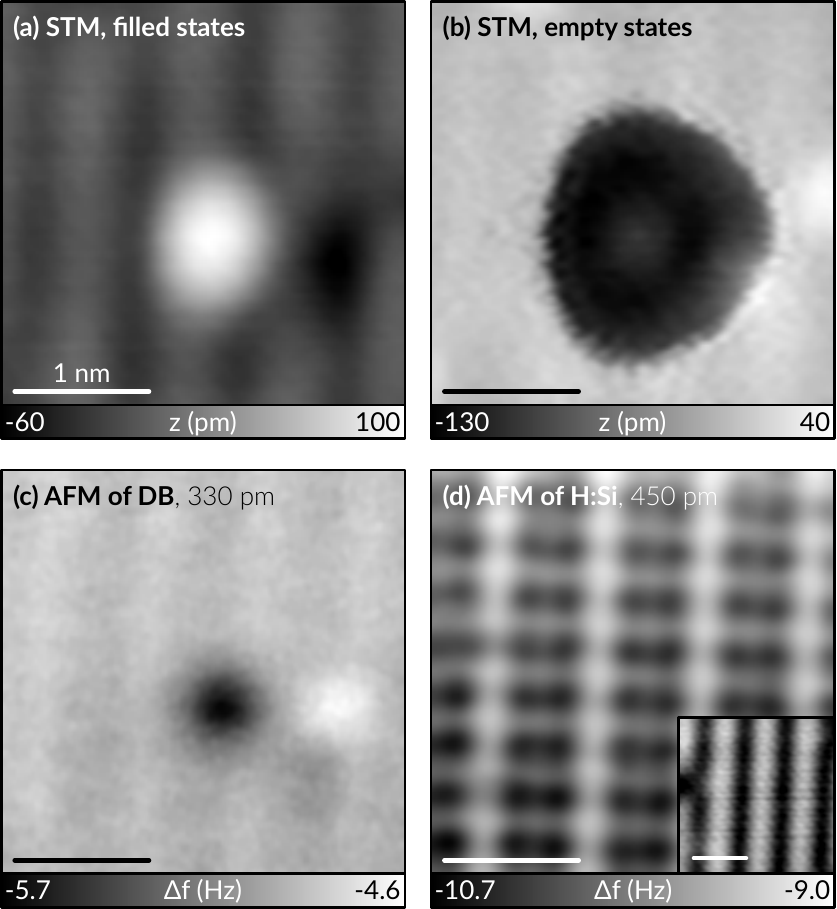}%
	\caption{\label{SiTip} STM/AFM imaging with a Si tip.
	(a) and (b) show filled and empty state STM images, respectively.
	(c) shows an AFM image of the same area as (a) and (b) at $\Delta z_{f} = 330\,\text{pm}$. 
	(d) shows an AFM image of a defect-free area at $\Delta z_{f} = 450\,\text{pm}$. Inset in (d) shows corresponding filled state STM imaging.
	All images were obtained with the same Si tip.
	%and its Laplace-filtered version, respectively.
	%Individual H:Si sites and the DB are atomically resolved by AFM. Note the \emph{dark} contrast of both the H-terminated Si dimer rows (c) and H-terminated Si adatoms (d) obtained by AFM with Si tips. 
	%At the reduced tip sample distance used in (d) imaging became unstable when scanning across the DB with the reactive Si tip resulting in rearrangement of the Si cluster at the tip apex showing indistinct contrast in subsequent AFM images.
	}%
\end{figure}

\begin{figure}
	\includegraphics{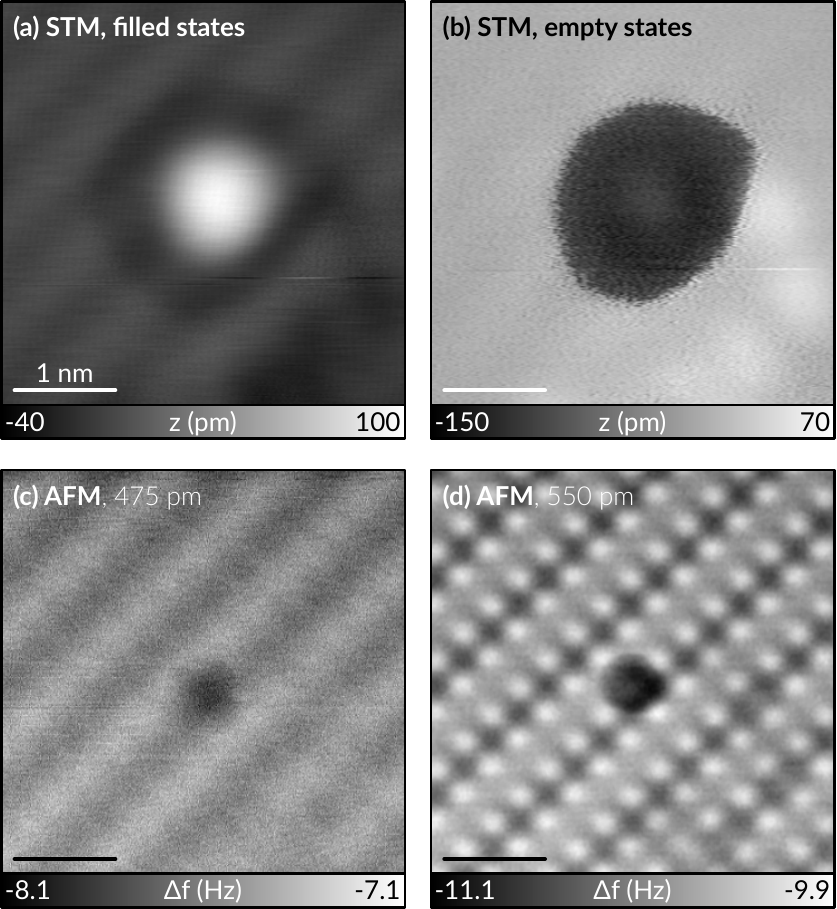}%
	\caption{\label{HTip} STM/AFM imaging with a H tip.
	(a) and (b) show filled and empty state STM images, respectively.
	(c) and (d) show AFM images at $\Delta z_{e} = 475\,\text{pm}$ and $\Delta z_{e} = 550\,\text{pm}$, respectively.
	All images show the same area and were obtained with the same H tip.
	%Individual H:Si sites, DBs and dihydrides are atomically resolved by AFM. Note the bright contrast of both the H-terminated Si dimer rows (c) and H-terminated Si adatoms (d) obtained by AFM with H tips. 
	}%
\end{figure}

All experiments were carried out at a temperature of $T = 5\,\text{K}$ using a qPlus sensor \cite{Giessibl1998} (eigenfrequency $f_0 = 25.066\,\text{kHz}$, stiffness $k \approx 1.8\,\text{kN/m}$, quality factor $Q \sim 10^5$) operated in frequency-modulation mode\cite{Albrecht1991}. The oscillation amplitude was set to $A = 150\,\text{pm}$. The bias voltage $V$ was applied to the sample. The PtIr tip of the sensor had previously been cut by focused ion beam and then, in UHV at low temperature, indented into a Cu(111) sample to obtain an atomically sharp, metallic, Cu covered tip. %For details of the AFM and sensor preparation see Refs.~\onlinecite{Gross2009,Gross2009b}.

\begin{figure*}
	\includegraphics{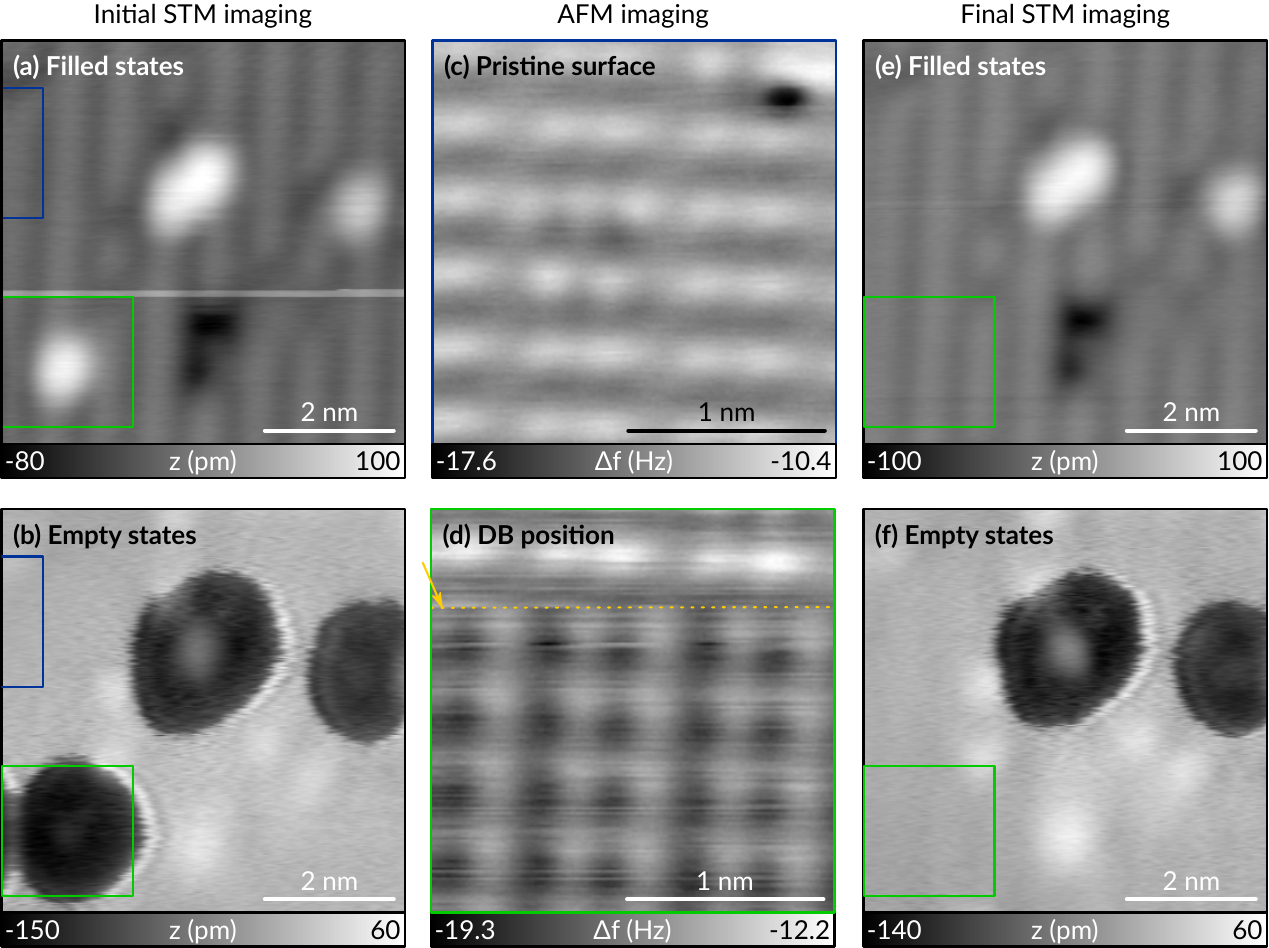}%
	\caption{\label{Healing} Hydrogenation of DB with a H tip.
	Filled (a) and empty (b) state STM images show several DBs recognized by characteristic dark halos in (b).
	(c) Constant-height AFM image ($\Delta z_{f} = 660\,\text{pm}$) of a small area of the H--Si surface [indicated by the partly visible blue square in panels (a-b)].
	(d) Constant-height AFM image ($\Delta z_{e} = 560\,\text{pm}$), zooming into the area indicated by a green square in the other panels. The slow scan direction of the image is from top to bottom. In the top part of the image the typical AFM contrast of H tips (hydrogenated Si sites appear bright, cf. Fig.~\ref{HTip}d) is still observed. At the position indicated by the orange dashed line and arrow, the image contrast suddenly changed to that of a Si tip (hydrogenated Si atoms appear dark, cf. Fig.~\ref{SiTip}d).
	After recording panel (d), filled (e) and empty (f) state STM images of the same area as shown in (a,b) show no trace of the former DB within the area of the green square indicating a locally defect-free H-terminated Si surface.
	}%
\end{figure*}

STM (AFM) images were recorded in constant-current (constant-height) mode and show the topography $z$ (frequency shift $\Delta f$). All filled (empty) state STM images shown here were acquired at $V = -2\,\text{V}$ ($V = 2\,\text{V}$) with the tunneling current set to $I = 2\,\text{pA}$. AFM images were recorded at $V = 0\,\text{V}$. The height offset, $\Delta z_{f}$ ($\Delta z_{e}$), is given for each AFM image with respect to the STM set point for filled (empty) state imaging above the bare H--Si surface. Both Si and H tips were created by bringing the tip into contact with the H--Si surface at $V = 2\,\text{V}$. The tips were assigned by their AFM contrast as explained in the following.

\section{Results}

STM and AFM images with the different tip functionalizations are shown in Fig.~\ref{SiTip} for the reactive Si tip and in Fig.~\ref{HTip} for the inert H tip, respectively. Our results are in agreement with previous studies \cite{Livadaru2011,Yurtsever2013,Taucer2014,Rashidi2016,Sharp2012,Sweetman2016,Labidi2017}. For STM images we observed a qualitatively similar contrast for both tips and the tips cannot unambiguously be assigned by STM images. At filled state imaging with both tips [Fig.~\ref{SiTip}(a) and Fig.~\ref{HTip}(a)] the dimer rows appear as faint protrusions and the DB appears as a protrusion with an lateral extension of about $1\,\text{nm}$ and an apparent height of about $70\,\text{pm}$. At empty state imaging [Fig.~\ref{SiTip}(b) and Fig.~\ref{HTip}(b)] the DB appears as a dark disc (depression) of about $2\,\text{nm}$ lateral extensions and an apparent depth of about $150\,\text{pm}$ with a faint protrusion in the center. Recently, the STM contrast could be successfully explained in detail by Refs.~\onlinecite{Livadaru2011,Taucer2014,Rashidi2016}. Importantly, the charge state of the DB is affected during imaging by both tip-induced band bending (TIBB) and, in the opposite way, by dynamic charging through electron attachment/detachment by tunneling from/onto the tip. Therefore, the contrast crucially depends on the applied voltage, the lateral and vertical tip location, and the doping of the substrate. Sharp transitions in the STM images indicate a change of the DB charge state as a function of lateral tip position, e.\,g.  observed at the rim of the dark disc in Fig.~\ref{SiTip}(b) and Fig.~\ref{HTip}(b).

We obtained constant-height AFM images with both tips at different tip heights $\Delta z$. To achieve atomic resolution in AFM, the tip is approached considerably compared to STM imaging. First, we discuss the frequency shift contrast of a Si tip. Fig.~\ref{SiTip}(c) shows an AFM image of the same area as Figs.~\ref{SiTip}(a,b) at $\Delta z_{f} = 330\,\text{pm}$ with the dimer rows appearing slightly darker (more negative $\Delta f$) than the background, and the DB appearing as a dark circular feature. Typically, approaching a Si tip closer to a DB results in tip changes and local destruction of the surface, presumably due to bond formation between tip and sample. %\footnote{see supplementary materials for details}. 
Therefore, to illustrate the atomic contrast of Si tips, Fig.~\ref{SiTip}(d) shows an AFM image of a defect-free area with the same Si tip at $\Delta z_{f} = 450\,\text{pm}$. The hydrogenated Si atoms appear as \emph{dark} circular features revealing the $2\times1$ reconstruction.

Next, we discuss the frequency shift contrast with a H tip. Figs.~\ref{HTip}(c) and \ref{HTip}(d) show AFM images of the same area as Figs.~\ref{HTip}(a,b) at $\Delta z_{e} = 475\,\text{pm}$ and $\Delta z_{e} = 550\,\text{pm}$, respectively. As in the case of Si tips described in the previous section, the dimer rows appear slightly darker than the background in Fig.~\ref{HTip}(c), and the DB appears as the darkest feature.
%Note that the darker contrast of the dimer rows has not been reported in Ref.~\onlinecite{Labidi2017}. We attribute this finding to (i) different dopant concentration of the surfaces resulting in different magnitude of TIBB, (ii) different contact potential of the (macroscopic) tips, (iii) the different tip generation procedures used, or a combination of any of these reasons. 
Importantly, in contrast to the Si tip, the hydrogenated Si atoms appear as \emph{bright} circular features with the H tip at closer approach [Fig.~\ref{HTip}(d)] in agreement with Ref.~\onlinecite{Labidi2017}. Dark corresponds to a decreased (more negative) frequency shift and, as we work in the attractive regime, i.\,e., on the rising branch of $\Delta f(z)$ (where a larger $z$ corresponds to a larger tip-sample distance), this corresponds to increased attractive forces. As described in Refs.~\onlinecite{Sharp2012,Sweetman2016} the Si tip yields an increase in attractive forces above hydrogenated Si sites while the passivated H tip yields a slightly reduced attraction above hydrogenated Si sites compared to the average H--Si surface. %The total forces are significantly larger for the reactive Si tip compared to the H tip, as seen in the $\Delta f$ scales of Fig.~\ref{SiTip}(d) and Fig.~\ref{HTip}(d).

The DB itself is imaged dark with both tips [Fig.~\ref{SiTip}(c) and Fig.~\ref{HTip}(c,d)]. The contrast can be tentatively explained by attractive electrostatic interaction due to a negatively charged DB. Note that the contrast changes for differently charged DBs. %, as demonstrated by Supplementary Fig.~5 (Ref.~\onlinecite{Note1}).
However, a detailed investigation of the charge state of the DB within the band gap, based on Kelvin probe force microscopy, goes beyond the scope of this paper.

We used Si tips to write DBs. Typically, a voltage of $V = 2.6\,\text{V}$ at tunneling currents on the order of $50\,\text{pA}$ was used to create an individual DB within less than $10\,\text{s}$. %\cite{Note1}
We did not optimize the writing of DBs in our work, which had been investigated systematically before \cite{Ballard2013,Kolmer2014,Moller2017,Wolkow2014}.

%(short paragraph on DB creation, can also go above to the experiment section, should be in). Do we also have an example of writing after erasing? Comment: There is missing something about reproducibility and yield. We should write, why we do not present complex structures. As a referee I would ask that. What is our restriction?

In the following, the controlled specific hydrogenation of a DB is demonstrated in Fig.~\ref{Healing}. First, we imaged an area with several DBs using STM with a H tip, see Fig.~\ref{Healing}(a,b). The H functionalization of the tip is confirmed by its characteristic AFM contrast (hydrogenated Si sites appear bright) on a DB-free area, as presented in Fig.~\ref{Healing}(c). A zoom-in on the DB in the region indicated by the green square is shown in Fig.~\ref{Healing}(d). During that AFM image [Fig.~\ref{Healing}(d)], which was recorded at a tip height closer to the surface as compared to Fig.~\ref{HTip}(d), the contrast suddenly changed from that of a H tip (hydrogenated Si sites bright) to the typical contrast of a Si tip (hydrogenated Si sites dark). After that AFM image no trace of the DB could be observed in subsequent STM images [Fig.~\ref{Healing}(e,f)] signaling that the DB was removed. We can conclude that the H atom of the H tip apex was transferred to the DB during the AFM image in Fig.~\ref{Healing}(d). In contrast to the hydrogen desorption process (DB creation) no voltage or tunnel current was applied for the DB hydrogenation.

\begin{figure}
	\includegraphics{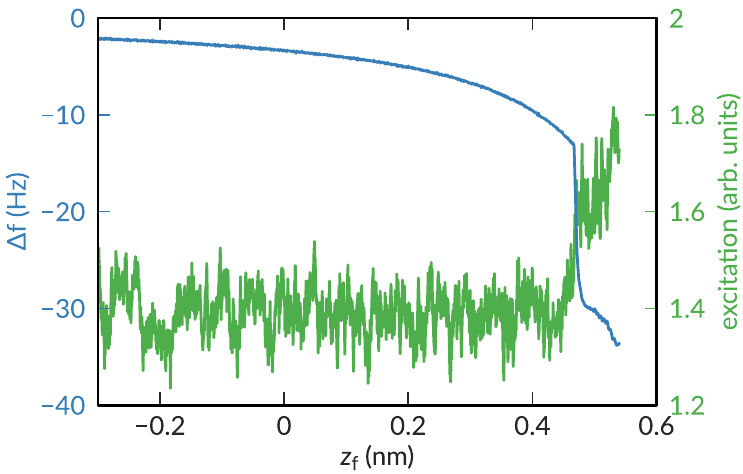}%
	\caption{\label{Erase2} DB passivation by vertical manipulation.
	A sharp change in the frequency shift $\Delta f$ vs. distance $z_\text{f}$ curve occurrs at $z_\text{f} \approx 0.45\,\text{nm}$ signalling DB passivation. Manipulation was performed at $V = 0\,\text{V}$.
	}%
\end{figure}

H passivation of DB was achieved routinely when a H tip was used. The H was transferred from the tip to the DB during imaging. Passivation could also be achieved by vertical manipulation, i.\,e. approaching a H tip vertically towards a DB as demonstrated in Fig.~\ref{Erase2}. In that case, DB passivation is signalled by a sudden change in frequency shift (to more negative values) and a correlating increase in the excitation signal of the sensor. In general, the requirement for DB passivation was a tip approach closer to the surface than necessary for typical AFM imaging. %In addition, it was also observed during STM imaging in a case where the H tip apex had been rearranged in a previous AFM image \cite{Note1}.
When the DB got passivated, the H tip was converted into a Si tip.

\section{Conclusion}
We demonstrated that an individual, deliberately chosen DB defect on a H--Si surface can be hydrogen-passivated to reestablish a locally defect-free surface. Passivation is achieved by atomic manipulation, transferring the H atom of a hydrogen-terminated tip into the DB. This method provides a proof of concept for error correction for hydrogen lithography and thus could significantly increase the yield for atomic precise structures written by hydrogen lithography. While AFM is very useful for the assignment of the tips, the correction method could also be established by STM only. %Moreover, it might be possible to apply the method also at room temperature.

\begin{acknowledgments}
We thank A. Fuhrer for discussions and help with the preparation of clean Si surfaces, and Rolf Allenspach for valuable comments on the manuscript.
The research leading to these results received funding from the ERC Advanced Grant CEMAS (agreement no. 291194), the ERC Consolidator Grant AMSEL (682144), and the EU project PAMS (610446).
\end{acknowledgments}

%\textsf{\color{red} Maximum length excluding references 4 pages!}

\end{document}